\documentclass{aa}

\usepackage{graphicx}
\usepackage[varg]{txfonts}
\usepackage{url}
\usepackage{natbib}

\begin{document}

\title{Evolution of redback radio pulsars in globular clusters}

\author{O. G. Benvenuto\inst{\ref{inst1},\ref{inst2}}
\and
M. A. De Vito\inst{\ref{inst1},\ref{inst2}}
\and
J. E. Horvath\inst{\ref{inst3}}}

\institute{Facultad de Ciencias Astron\'omicas y Geof\'{\i}sicas, Universidad
Nacional de La Plata, Paseo del Bosque S/N (B1900FWA), La Plata, Argentina
\label{inst1}
\and
Instituto de Astrof\'{\i}sica de La Plata (IALP),
CCT-CONICET-UNLP. Paseo del Bosque S/N (B1900FWA), La Plata, Argentina
\label{inst2}
\and
Instituto de Astronomia, Geof\'{\i}sica e Ci\^encias Atmosf\'ericas,
Universidade de S\~ao Paulo, R. do Mat\~ao 1226 (05508-090),
Cidade Universit\'aria, S\~ao Paulo SP, Brazil
\label{inst3}
}

\date{Recived / Accepted}

\abstract{We study the evolution of close binary systems composed of a
normal, intermediate mass star and a neutron star considering a chemical
composition typical of that present in globular clusters (Z = 0.001).} 
{We look for similarities and differences with respect to solar
composition  donor stars, which we have extensively studied in the past. As a
definite  example, we perform an application on one of the redbacks located in a
globular  cluster.} 
{We performed a detailed grid of models in order to find systems that represent 
the so-called redback binary radio pulsar systems with donor star masses
between 0.6 and 2.0 solar masses and orbital periods in the range 0.2 - 0.9
days.} 
{We find that the evolution of these binary systems is rather similar to those
corresponding to solar composition objects, allowing us to account for
the  occurrence of redbacks in globular clusters, as the main physical
ingredient is the irradiation feedback. Redback systems are in the quasi-RLOF
state,  that is, almost filling their corresponding Roche lobe. During the
irradiation cycle the system alternates between semi-detached and detached
states. While detached the system appears as a binary millisecond pulsar, called a redback.
Circumstellar material, as seen in redbacks, is left behind after the previous
semi-detached phase.} 
{The evolution of binary radio pulsar systems considering irradiation 
successfully accounts for, and provides a way for, the occurrence of redback  pulsars
in low-metallicity environments such as globular clusters. This is the case despite possible
effects of the low metal content of the donor star that could drive systems
away from redback configuration.}

\maketitle

\section{Introduction}
\label{sec:introd}

Redback pulsars represent a recently identified and fast growing family of
binary,  eclipsing radio pulsars whose companion is a low-mass star $0.1
\lesssim M_{2}/M_{\odot} \lesssim 0.7$ (being $M_2$ the mass of the companion)
on almost circular orbits with periods  $0.1 \lesssim P_{\rm orb}/d \lesssim
1.0$.  Black widows form another family of pulsars with  orbital periods in the
same range but with companion stars that are appreciably  lighter, $M_{2}
\lesssim 0.05 M_{\odot}$. Among detected redbacks, some belong to the Galactic
disk population whereas others reside in globular clusters (GCs). A recent
listing of these  system can be found in  A. Patruno's
catalogue\footnote{\url{www.apatruno.wordpress.com/about/millisecond-pulsar-catalogue/}
\label{footnote:patruno} }.

At present there are different proposals concerning the formation of redback
pulsars.  \cite{2013ApJ...775...27C} studied close binary systems with
evaporation, considering  different efficiencies due to geometrical effects.
These authors concluded that redback and  black widow pulsars follow different
evolutionary paths depending on the strength of evaporation, finding that black
widows do not descend from redback pulsars.  \cite{2015MNRAS.446.2540S} proposed
that redbacks are formed via accretion induced  collapse of
oxygen-neon-magnesium white dwarfs (WDs). The neutron star (NS) becomes a pulsar
whose irradiation evaporates the companion star. They considered that pulsar
emission inhibits further accretion, finding that the whole redback region of
the plane $M_{2}-P_{\rm orb}$, where redbacks are located, can be populated in
this way. \cite{2014ApJ...786L...7B} (BDVH14) considered the evolution of close
binary systems (CBSs) including the effects of evaporation and irradiation
feedback.  These authors found that mass transfer  occurs in several cycles when
the system has an orbital period and donor mass in the range corresponding to
redbacks. They  concluded that black widows descend from redbacks.  In this
paper we explore this evolutionary scenario further. 

Irradiation feedback \citep{2004A&A...423..281B} is an important phenomenon for
a detailed model of mass transfer episodes in CBSs containing a NS. When the
donor star fills its Roche lobe it begins to transfer mass to the NS, which may
accrete a fraction of this mass.  Because matter falls down the NS gravitational
well, it releases X-rays, irradiating the donor star.  The irradiated part of
the donor star surface becomes partially blocked to release energy emerging
from  its interior. This effect may make the star undergo a number of mass
transfer cycles. This is in sharp contrast with the classical results of
standard CBS evolution, which predict long standing, continuous mass transfer
stages. This irradiation model may be not sophisticated enough to explain,  for
example, the light curve observed for the companion of PSR~J1544+4937  (see e.g.
\citealt{2014ApJ...791L...5T}). Nevertheless, this model enables a comprehensive
description of the irradiation feedback and its implication for BWs, which we
have used in this work.

Later on, the evaporation of the donor star driven by the pulsar wind becomes
important.  This is strongly suggested by the observations of the famous
eclipsing radio pulsar PSR~1957+20, which is the prototype of the family of 
black widow pulsars \citep{1988Natur.333..832P}.  This effect makes the CBS
undergo strong mass and angular  momentum losses, allowing models to reach
orbital periods of days when the leftover donor star achieves very small mass.

We (BDVH14) presented several evolutionary tracks crossing the redback region
of the $M_{2}-P_{\rm orb}$ plane. Most of these undergo several mass transfer
cycles.  Typically CBSs undergo mass transfer during a small fraction of each
cycle and otherwise  remain detached. In the detached stage it should be
possible to  observe the CBS as a binary radio pulsar, while during mass
transfer it may be detected as a low-mass X-ray binary system (LMXB).

During mass transfer it may also be possible to detect the radio pulsar if the
accretion disk surrounding the NS suffers some instability, as is strongly
suggested by some recent key observations. PSR~J1023+0038 is a Galactic field
redback that is  detected undergoing a transition from millisecond pulsar
(MSP) to LMXB \citep{2014ApJ...790...39S}. PSR~J1824-2452I is a redback in a
GC switching between accreting millisecond X-ray pulsar and redback states
\citep{2013Natur.501..517P}. During the fraction of the  cycles in which the
pair is detached, the donor star is very similar in size to its corresponding
Roche lobe. This is in the {\it quasi-Roche} lobe overflow state (quasi-RLOF).
This feature provides an interpretation \citep{2015ApJ...798...44B} of the
recent observations of PSR~J1723-2837 presented by \citet{2013ApJ...776...20C}
and is suggested to be important for the characterization of the redback
systems.

If the radio pulsar becomes observable, thereby requiring the dissipation of
the innermost part of the accretion disk, likely by some instability
mechanism, the system must be surrounded by the material transferred by the
donor star, which is in the RLOF state. Otherwise, if the system is in the
quasi-RLOF,  its neighbourhood should be populated by some material
evaporated from the  donor star (see Fig.~2 of \citealt{2012ApJ...753L..33B},
in particular inset  A). For a given cycle, the evaporation rate is far weaker
than the  mass transfer rate at RLOF. Thus, compared to the corresponding
Roche  lobe, the eclipsing region of systems in the quasi-RLOF state should
be  smaller than that corresponding to systems in the RLOF state. There are
good  examples of both situations: the Galactic PSR~J1723-2837 shows eclipses
from  material located in a region twice its Roche lobe, whereas 
PSR~J1824-2452H, belonging to the GC M28, shows eclipses from a
region six times its Roche lobe.

All the theoretical results referred to above correspond to objects with solar
composition. Because a large percentage of the detected redbacks belong to GCs
(at present eight systems are located in the Galactic field and another
ten are found in GCs; see Table~\ref{table:RBs-in-GCs}), it is natural to
extend our  theoretical results to the case of stars with a lower metal content,
which is typical of Population II. This is because for low metallicity the whole
picture has to be checked. A failure to produce intermittent mass transfer and
settle into the quasi-RLOF state would represent a challenge  to the proposed
interpretation. 

\begin{table*}
 \caption{Redbacks in GCs from the Patruno catalogue. The minimum 
estimated mass  and the association are taken from ATNF data base. In the
case of (*), the estimated mass of the companion is that of
\citet{2003A&A...397..237O}. The metal content for the association to 
which the system belongs are taken from the references  quoted at the end of 
the Table.}
 \label{table:RBs-in-GCs}
 \centering
 \begin{tabular}{lccc}
 \hline\hline
 PSR         & Minimum Mass [$M_{\odot}$] & Association & ${\rm [Fe/H]}$ \\
 \hline
 J0024-7204W & 0.123977 & 47Tuc & $-0.75$ (a) \\
 J1701-3006B & 0.121576 & M62, NGC6266 & $-1.15 \pm 0.02$ (b)\\
 J1740-5340 (*)  & 0.30 & NGC6397 &  $-2.28 \pm 0.01$ (c) \\
 J1748-2021D & 0.121182 & NGC6440 & $-0.56 \pm 0.02$ (d)\\
 J1748-2446A & 0.087283 & Ter5    & $-0.21$ (e) \\
 J1748-2446ad& 0.138630 & Ter5    & $-0.21$ \\
 J1748-2446P & 0.367339 & Ter5    & $-0.21$ \\
 J1824-2452H & 0.169625 & M28, NGC6626& $-1.16 \pm 0.2$ (f)\\
 J1824-2452I & 0.174484 & M28 ,NGC6626& $-1.16 \pm 0.2$ \\
 J2140-2310A & 0.098838 & M30 & $-2.01 \pm 0.06$ (g)\\
 \hline
\multicolumn{4}{l}{ }\\
\multicolumn{4}{l}{ (a) \citet{2014MNRAS.437.3274V}}\\
\multicolumn{4}{l}{ (b) \citet{2014MNRAS.439.2638Y}}\\
\multicolumn{4}{l}{ (c) \citet{2012ApJ...754...91L}}\\
\multicolumn{4}{l}{ (d) \citet{2008MNRAS.388.1419O}}\\
\multicolumn{4}{l}{ (e) \citet{2007A&A...470.1043O}}\\
\multicolumn{4}{l}{ (f) \citet{1985ApJ...298..572S}}\\
\multicolumn{4}{l}{ (g) \citet{2013A&A...555A..36K}}\\
\multicolumn{4}{l}{ }\\
 \hline 
 \end{tabular}
\end{table*}

The response of irradiated donor stars in binary systems is largely 
determined by the occurrence of an outer convective zone (OCZ). It is well
known  that for a given mass and evolutionary state, Population II stars are 
more compact than those of solar composition. It is also well known that,
owing  to the low metal content of the mixture, at temperatures typical of
outer  stellar layers, the material has a lower opacity, which in turn,
directly  affects the behaviour of the OCZ. Furthermore, as stated above, many
redbacks belong to GCs. In this context, a  detailed exploration of the whole
evolution of these systems is in order.  We show below that the very existence
of redbacks in GCs can be accounted for by irradiated models, as for solar
composition objects.  The remainder of this work is organized as follows: In
Section~\ref{sec:resul} we present  our numerical results.   We apply them in
Section~\ref{sec:aplicando} to the case of PSR~J1824-2452H. Finally, in
Section~\ref{sec:discu_conclu} we discuss the meaning of these results and
give some concluding remarks.

\section{Results} \label{sec:resul}

We employed our stellar binary code originally described by
\citet{2003MNRAS.342...50B} with the modifications cited by BDVH14,
considering irradiation feedback on to the donor star without any sensible
screening by the  accretion disk itself (see e.g.
\citealt{2001NewAR..45..449L}) and neglecting radio ejection effects
(\citealt{1989ApJ...336..507R}; \citealt{2001ApJ...560L..71B}).  We considered
that evaporation is not relevant for the evolutionary stages we are
interested in here. In addition, in the computation of the stellar models, we
neglected the effects of rotation. Other illuminating models, of an indirect
type, have been considered in the literature \citep{2016arXiv160603518R}
motivated by high-resolution observations and these may be useful  towards a 
compelling modelling of these systems. This possibility is not considered
here  and will be addressed in the future.

As BDVH14, we considered several initial masses and orbital periods
considering that the NS is able to accrete a fraction $\beta$ of the mass
transferred by the donor star, assuming that the absolute upper limit is given
by the Eddington accretion rate, $\dot{M}_{\rm Edd}= 2\times 10^{-8}$
$M_{\odot}\,yr^{-1}$.  For a given set of initial conditions ($M_2$, $P_{\rm
orb}$) we considered four different sequences with and without irradiation
feedback. The accretion luminosity released by the NS is  $L_{\rm acc}= G
M_{\rm ns} \dot{M}_{\rm ns}/R_{\rm ns}$, where $M_{\rm ns}$,  $R_{\rm ns}$,
and $\dot{M}_{\rm ns}$ are the mass, radius, and accretion rate of  the NS,
respectively, and $G$ is the Newton gravitational constant. Assuming
isotropy,  the energy flux that falls onto the donor star and
effectively participates in  the irradiation feedback  process is $F_{\rm
irr}= \alpha_{\rm irrad} L_{\rm  acc}/4\pi a^2$. Here $a$ is the orbital
radius and $\alpha_{\rm irrad}$,  considered a free parameter, is the fraction
of the incident flux that  effectively participates in the feedback 
process. We considered $\alpha_{\rm irrad}= 0.00, 0.01, 0.10,$ and $1.00$. 

Specifically, we considered an initial canonical mass of 1.4~$M_{\odot}$ for
the NS, while donor star masses $M_{2}$ were considered from 0.6~$M_{\odot}$
to 2.0~$M_{\odot}$ with steps of 0.1~$M_{\odot}$. While it is well
known that there is a wide distribution for observed NS masses, the results
presented in this paper are not sensitive to this mass. The initial orbital
periods were set from  0.2$\,$d to 0.9$\,$d, which is appreciably shorter
than those relevant for solar composition objects because low-metallicity
stars are smaller and therefore they should begin mass transfer at a closer
separation. If the initial periods were longer, the systems would {\it not}
evolve towards the region of the $M_{2}-P_{\rm orb}$ plane of interest. In
Table~\ref{table:RBs-in-GCs}, we show ${\rm [Fe/H]}$ for the GCs that host
redback systems.  The relation between the ratio ${\rm [Fe/H]}$ and the
metal content $Z$ is  given by ${\rm log_{10} Z = [Fe/H]} - 1.7$. We
performed a fitting to ${\rm [Fe/H]}$ and $Z$ taken from 
\url{www.galev.org}  \citep{2009MNRAS.396..462K}. Therefore, for the GCs
listed in Table~\ref{table:RBs-in-GCs},  $Z$ lies between 0.0001 and 0.012.
In our calculations, we used a fixed metallicity of  Z = 0.001 that is
representative of GCs. We do not aim in this work to perform a detailed
fitting of a particular redback system in one of these GCs, but rather to
explain their existence in general.  We evolved these CBSs throughout the
redback region. 

\begin{figure*}
 \centering
 \includegraphics[width=12cm, angle=270]{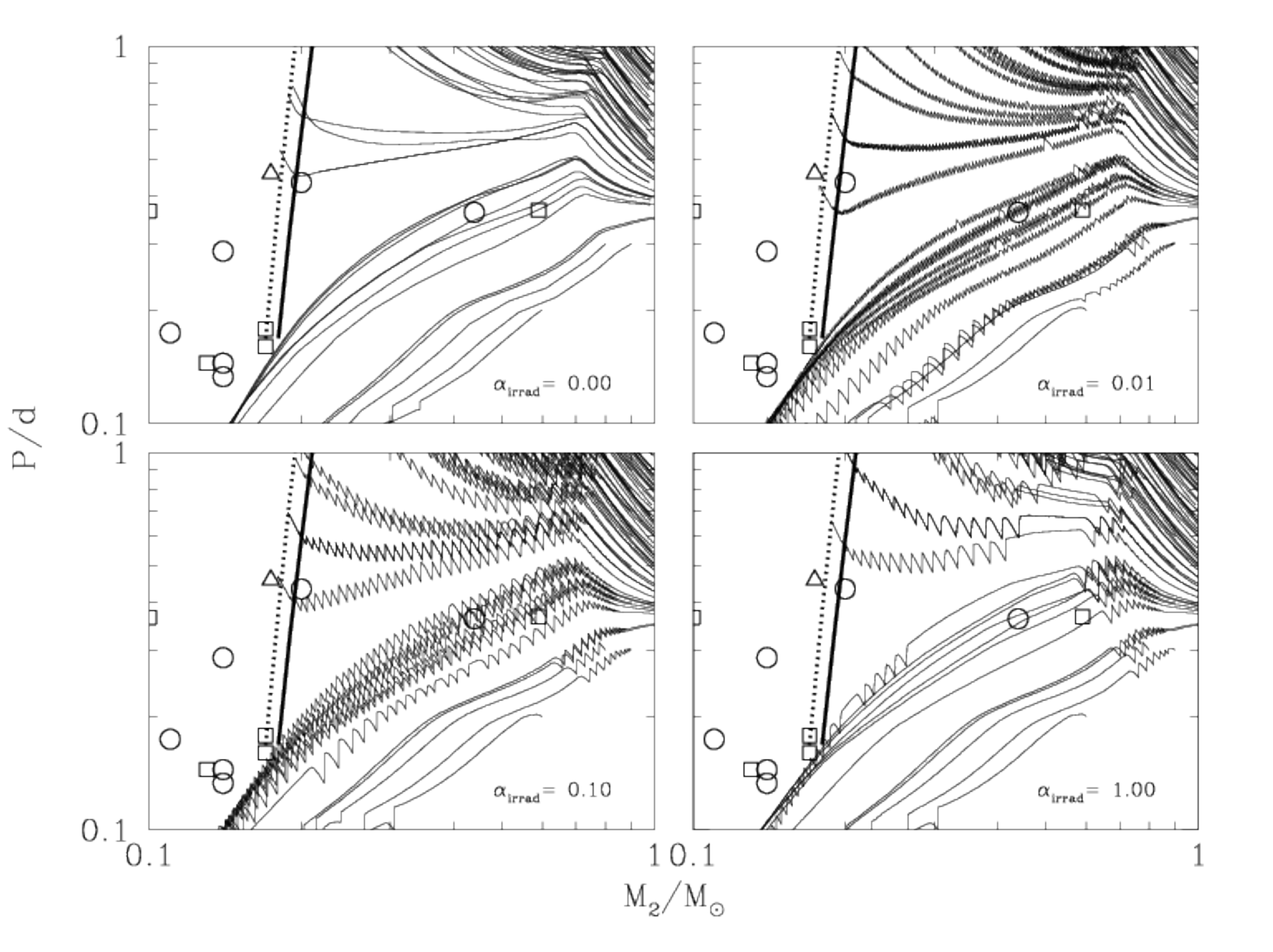}
  \caption{Donor mass vs. orbital period for CBSs, for the case of a typical
GC chemical composition ($Z= 0.001$) in the region where redbacks are located
together with the corresponding observational data. Symbols denote {\it
minimum} masses for radio pulsar companions. Circles and squares represent
redbacks and accreting MSPs, whereas the triangle denotes the case of a system
that switches between these states, all from the Patruno'
catalogue$^{\ref{footnote:patruno}}$. The thick solid (dotted) line corresponds to
the mass radius relationship expected for Population II (I) composition
objects given by \citet{1999A&A...350..928T}.}
  \label{Fig:mass_vs_peri}
\end{figure*}

\begin{figure*}
\centering
 \includegraphics[width=12cm]{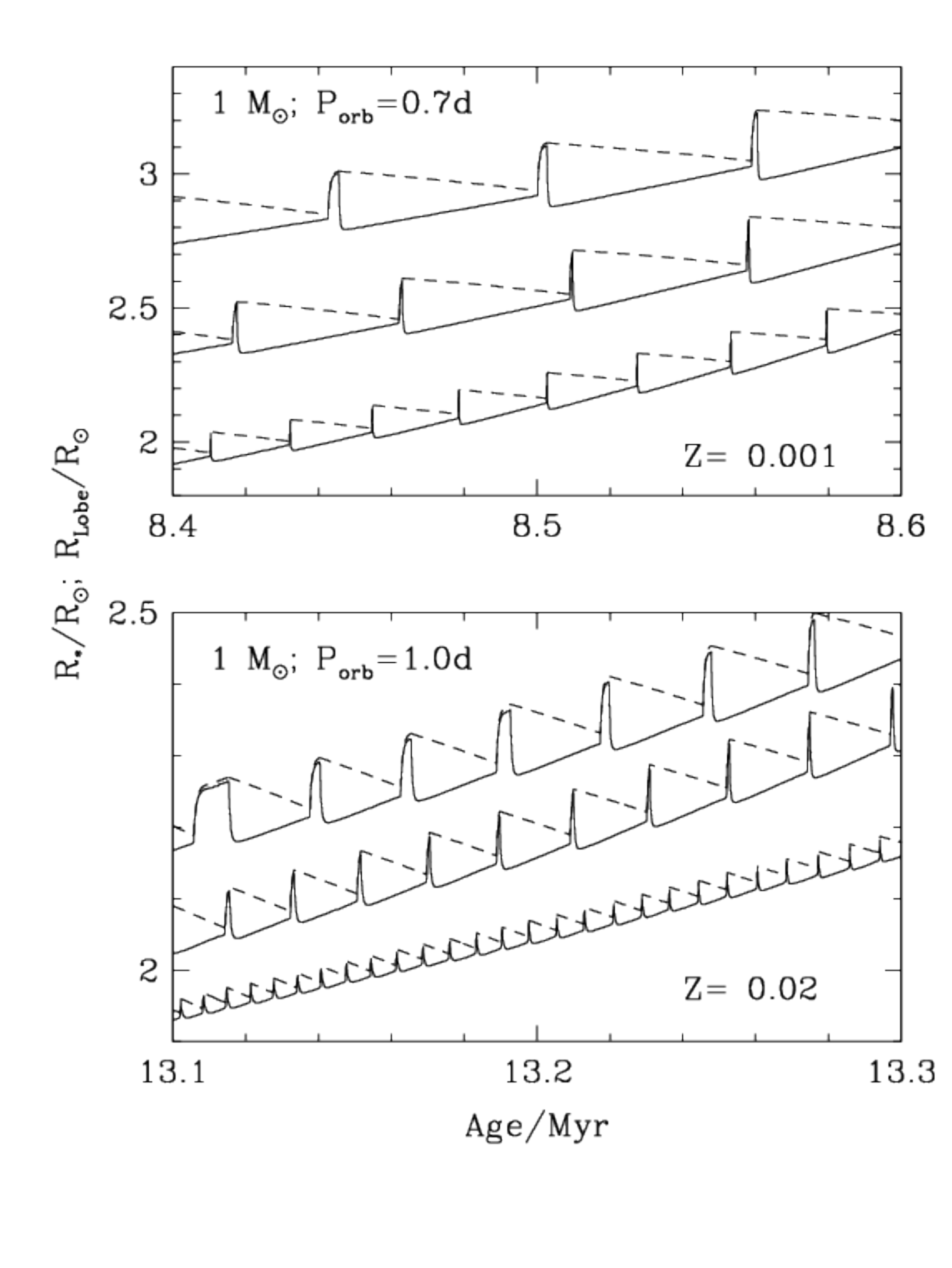}  
  \caption{Evolution of the stellar radius (solid lines) and the corresponding
equivalent Roche lobe radius (dashed lines) for models with $Z= 0.020$ and $Z=
0.001$ (donor mass and orbital periods are indicated in each panel). For the
case of  $Z= 0.001$ radii corresponding to $\alpha_{\rm irrad}=0.10 (1.00)$ have
been subject to a vertical offset of 0.05 (0.20). For models with $Z= 0.020$ the
offset was $0.5$ and $1.0$, respectively.}
  \label{Fig:radios}
\end{figure*}

The main results of this work are presented in
Figs.~\ref{Fig:mass_vs_peri}-\ref{Fig:bichito}. Fig.~\ref{Fig:mass_vs_peri}
presents the donor mass versus orbital period relationship  corresponding to the
four considered cases of irradiation feedback, parametrized by $\alpha_{\rm
irrad}$.  We included the data taken from the Patruno
catalogue$^{\ref{footnote:patruno}}$, corresponding to redbacks, accreting X-ray
MSPs, and the system PSR~J1824-2452I that switches between these states, which are all
observed in GCs. The observations provide {\it minimum} masses for radio
pulsar companions.

If irradiation feedback is neglected ($\alpha_{\rm irrad}= 0.00$) the systems
undergo continuous RLOF.  On the other hand, if irradiation feedback is allowed
to operate (from $\alpha_{\rm irrad} \gtrsim 0.01$ on\footnote{Here we do {\it
not} try to determine the minimum critical value of $\alpha_{\rm irrad}$ for
cycles to occur.}), we find that the evolutionary tracks cover the same region
of the $M_{2}-P_{\rm orb}$ plane, but now undergo mass transfer cycles just as
in the case of solar composition objects. The occurrence of these cycles make
our models plausible to account for the detected redbacks belonging to GCs. 
However, these cycles do not reproduce the transitions from the state of LMXB to
redback, and conversely, that have been recently observed (PSR~J1023+0038,
PSR~J1825-2452I and XSS~J12270-4859). These would respond to dynamic processes
associated with the details of the accretion disk, with short timescales (of the
order of months - years).

In Fig.~\ref{Fig:radios} we compare the cyclic behaviour of a CBS with solar
composition ($M_{2}= 1 M_{\odot}$, $P_{\rm orb}= 1\,$d) to the case of another
CBS with $Z= 0.001$ ($M_{2}= 1 M_{\odot}$, $P_{\rm orb}= 0.7\,$d).  The
$P_{\rm orb}$ for the CBS with typical GC composition was set to a lower
value.  This has been carried out to obtain comparable behaviours because
donors with GC composition are smaller so they should be closer to the NS. In
Fig.~\ref{Fig:radios} we show the three $\alpha_{\rm irrad} > 0.00$ considered
in this paper, finding that cycles are in general longer, therefore the total
number of cycles is fewer compared to the case of solar composition models.
Again, analogous to the solar composition calculations, the cycles are longer
when the irradiation becomes stronger (larger $\alpha_{\rm irrad}$).

These results support the idea that GC redback evolution is indeed globally
similar to that of solar composition objects. In particular, redback
companions should be either in RLOF or quasi-RLOF conditions. This is in
agreement with the fact that donor stars nearly fill their corresponding Roche
lobe as discussed by \citet{2013IAUS..291..127R}. Moreover, some GC redbacks
should evolve to become black widows, while others should become low-mass
helium WDs (see BDVH14).

\section{Application to a specific case} \label{sec:aplicando}

There are various redback systems in GCs with optically detected companions. 
The companion of PSR~J1740-5340 in NGC~6397 is a bright and
tidally deformed star. The luminosity and colours for this object are
incompatible with a WD star \citep{2001ApJ...561L..93F}. PSR~J1701-3006B is in
the GC NGC~6266. Its companion has been detected by \citet{2008ApJ...679L.105C}.
Similar to the companion of PSR~J1740-5340, it has a luminosity that is comparable
with the turnoff of the GC but significantly reddened. Both companion stars
have high filling factors of their Roche lobe. Finally, the companion of
PSR~J0024-7204W was identified as a faint blue star with radio and optical
observations that are compatible with a main-sequence star
\citep{2002ApJ...579..741E}.

The companion of PSR~J1824-2452H was identified by \citet{2010ApJ...725.1165P}.
This system belongs to the GC M28. It has an orbital period of $P_{\rm orb} =
0.435\,$d and shows eclipses for about $20\%$ of its period. The companion
colours are $U = 21.99$, $V = 20.58$ and $I = 19.49$. It is about $1.5\,$mag
fainter than the turnoff of the cluster and slightly bluer than the 
main-sequence. This object is too red and bright to be compatible with a WD. Based on
models of \citet{2008A&A...482..883M}, for isolated stars with ${\rm
[Fe/H]}=-1.27$ that are appropriate for this cluster, and assuming a colour excess $E(B -
V) = 0.4$ and a distance modulus $(m - M)_V = 14.97$
\citep{1996AJ....112.1487H}, the authors estimate for the companion of
PSR~J1824-2452H a mass of $M_{2}=0.68~M_{\odot}$, a temperature of 6000$\,$K, and a
radius of $R_{2}=0.64~R_{\odot}$. However, the mass function derived from radio
observations ($f_1 = 0.00211277~M_{\odot}$), the light curve shape  and the
occurrence of eclipses suggest a high orbital plane inclination and, hence, an
even lower mass of the companion. Assuming  an inclination of $i = 60^{\circ}$
(the median for the inclination angles) and a NS mass of $M_{\rm ns} =
1.4~M_{\odot}$, the mass of the companion is just $M_{2}\approx 0.2~M_{\odot}$.
In fact, the authors perform an iterative procedure to obtain the inclination,
the mass ratio $Q = M_{\rm ns} / M_2$, and the Roche lobe filling factor. The
best-fit model leads to $i \simeq 65^{\circ}$, $Q \simeq 7$  and a  Roche lobe
filling factor of about 1. The size of the Roche lobe for this star is of $R_{\rm L}
\approx 0.65~R_{\odot}$. This accounts for the observed luminosity of the
companion of PSR~J1824-2452H  but seems too small to cause the observed radio
eclipses (corresponding to an eclipsing region of about $3.3~R_{\odot}$ in
size). This suggests that this material extends beyond the Roche lobe.

\begin{figure*}
\centering
 \includegraphics[width=12cm]{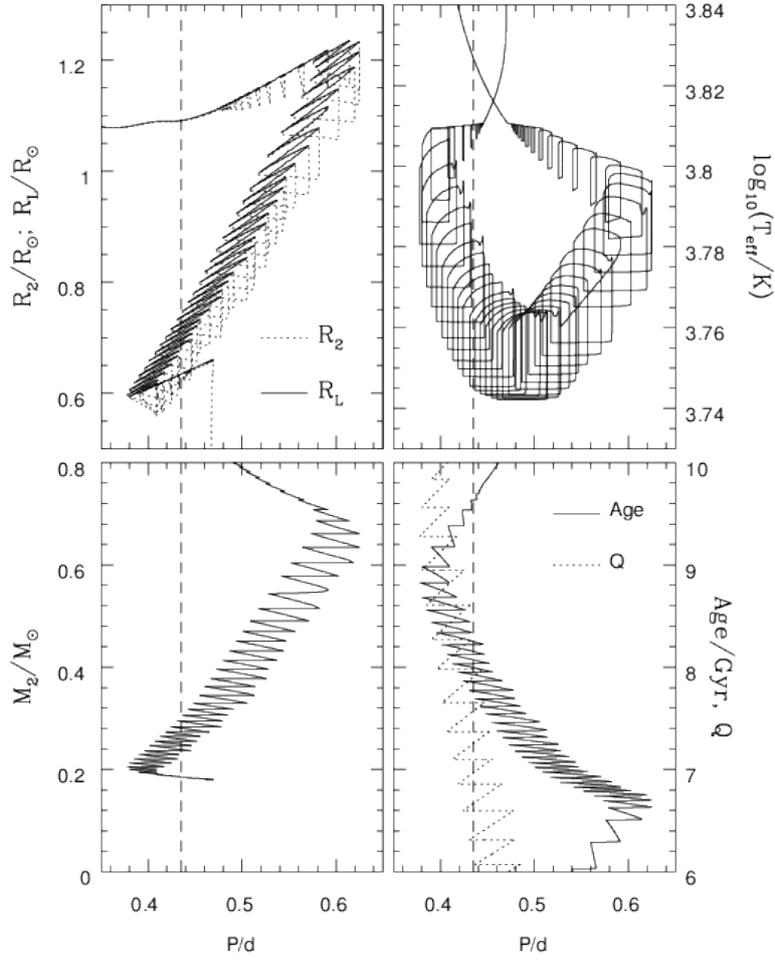}
  \caption{Evolution of a CBS with $Z= 0.001$, initial masses $M_{2}= 1.30
M_{\odot}$ and $M_{\rm ns}= 1.40 M_{\odot}$, $P_{\rm orb}=0.35\,$d, $\beta=0.50$, and
$\alpha_{\rm irrad}= 0.10$. All quantities are shown as a function of  $P_{\rm
orb}$.  Left upper panel shows the radius of the star and corresponding Roche
lobe (dotted  and solid lines, respectively);  right upper panel depicts the
effective temperature; left lower panel indicates the  donor mass and the right lower
panel shows the age and mass ratio (solid and dotted lines,  respectively). The
orbital period of PSR~J1824-2452H system, $P_{\rm orb}= 0.435\,$d, is  shown with
a vertical dashed line. The preferred $M_{2} \approx 0.2 M_{\odot}$,
$R_{2} \approx 0.65 R_{\odot}$, $\rm log_{10}{(T_{eff}/K)} \approx 3.78$, and  $Q
\approx 7$  given by \citet{2010ApJ...725.1165P} are consistent with these
results.}
  \label{Fig:bichito}
\end{figure*}

It is not the aim of this work to fit our models to the observed parameters of
all the binary systems enumerated above. However, we applied our calculations
to find a possible progenitor that accounts for the present state of one
representative system, namely PSR~J1824-2452H. An inspection of our models
yields that the CBSs with initial masses of $M_{2}= 1.30 M_{\odot}$ and
$M_{\rm ns}= 1.40 M_{\odot}$, $P_{\rm orb}=0.35\,$d, $\beta=0.50$, and
$\alpha_{\rm irrad}= 0.01, 0.10$ provide a suitable evolutionary frame to
characterize the PSR~J1824-2452H system. The results corresponding to these
evolutionary tracks are very similar to each other,  where the main difference
is the number of mass transfer cycles. Because the model with $\alpha_{\rm
irrad}= 0.10$ undergoes fewer cycles, it is easier to employ for the sake of
the discussion. Its main characteristics are presented in
Fig.~\ref{Fig:bichito}.  In this figure, we show the radii of the donor
$R_{2}$ and the Roche lobe $R_{\rm L}$ as a function $P_{\rm orb}$ (left upper
panel),  effective temperature $T_{\rm eff}$ (right upper panel), donor mass
$M_{2}$ (left lower panel), and the age and mass ratio $Q$ (right lower
panel). At the observed $P_{\rm orb}$ (indicated in Fig.~\ref{Fig:bichito}
with a vertical dashed line) we find good agreement between the observed
characteristics of the donor star \citep{2010ApJ...725.1165P} and our model at
an age of approximately 8 Gyr. Our theoretical age is shorter than the best
fit to the main-sequence of the host cluster (13$\,$Gyr;
\citealt{2010ApJ...725.1165P}). There are several parameters to be selected
for a CBS to study any observed object  that largely affect the time the
system spends to reach a given evolutionary stage; in this context the donor
mass and orbital period are especially relevant.  The exercise described above
to characterize the evolutionary status of PSR~J1824-2452H shows the
suitability of the models as a tool to study the observed redbacks in GCs.
However, it is not worthwhile to fit to the age of this particular system.

As stated above, PSR~J1824-2452H undergoes extended eclipses corresponding to
a region of about $3.3\ R_{\odot}$, far larger than its associated Roche lobe
and showing that the system is losing mass. Thus, this system should be in a
RLOF stage and the pulsar should be visible because of accretion disk
instabilities and/or radio ejection. Otherwise it would be seen as a
LMXB. Still, it may be argued that CBSs with the same characteristics
($M_{2}$, $M_{\rm ns}$, $P_{\rm orb}$) but with $\alpha_{\rm irrad}= 0.00$ and $1.00$,
which do not undergo mass transfer cycles for the same conditions
corresponding to PSR~J1824-2452H, may provide a suitable evolutionary frame.

\section{Discussion and conclusions} \label{sec:discu_conclu}

In this work we have presented calculations of the evolution of CBSs formed by
intermediate  mass donor stars together with a NS for the case of low
metallicity  ($Z=0.0010$), compatible with the abundances typically found in
GCs.  We considered standard CBS evolution and models in which irradiation
feedback is included.  We find that the global behaviour of CBSs is comparable
to that found by BDVH14 for the case of donor stars with solar composition. As
we discussed there, irradiation feedback gives rise to a cyclic mass
transfer behaviour, switching from detached to semi-detached conditions, for
which systems should be observable as radio pulsars and low-mass X-ray
binaries, respectively. As in the case of solar composition objects,  the
number of mass transfer cycles undergone by a given CBS increases when
$\alpha_{\rm irrad}$  decreases. However, in the case of low-metallicity
objects the number of cycles is remarkably lower than underwent by systems of
high metallicity for the same $\alpha_{\rm irrad}$. 

While our models  with irradiation feedback and evaporation provide a plausible
scenario  for the occurrence of redback pulsars, there is no reason to consider
it to be the only such scenario. Indeed, other very different proposals have been
presented as those of \cite{2013ApJ...775...27C} and \cite{2015MNRAS.446.2540S}.
These works have explored possible conditions  for the evolution of CBSs that is
different from those we considered (very efficient evaporation and  accretion
induced collapse of oxygen-neon-magnesium WDs). The differences between
these scenarios and ours are qualitative. While we found that black widows should
descend from redbacks, they claim that these  families of binary pulsars are due
to completely different evolutionary processes. It may be that all these ways of
forming redbacks can occur in nature; whether or not this is the case is still an
open problem.


We would like to thank our referee Chris Tout for his corrections and
suggestions, which have helped us to improve the original version of this
paper.


O.G.B. is member of the Carrera del Investigador Cient\'{\i}fico of the Comisi\'on de Investigaciones 
Cient\'{\i}ficas de la Provincia de Buenos Aires (CICPBA), Argentina. M.A.D.V. is member of the Carrera del 
Investigador Cient\'{\i}fico of the Consejo Nacional de Investigaciones Cient\'{\i}ficas y T\'ecnicas 
(CONICET), Argentina. J.E.H. has been supported by 
Fapesp (S\~ao Paulo, Brazil) under Grant No 2013/26258-4 and CNPq, Brazil
funding agencies.


\bibliographystyle{aa}


\bibliography{rb-glob-aa}

\end{document}